# Toward Architectural Knowledge Sustainability
## New Opportunities to Extend the Longevity of Systems

Rafael Capilla

Elisa Yumi Nakagawa

Uwe Zdun

Carlos Carrillo

Complex software systems must be maintained for years or decades, and the effort and cost to maintain them are often high, involving continuous refactoring to ensure their longevity in the face of changing requirements. In this article, we introduce the notion of architectural knowledge (AK) sustainability as a new concept to support architects dealing with the evolution of long-lived systems. Architecture sustainability refers to the ability of the architecture to endure over time with the minimum number of refactoring cycles possible. We suggest that sustainability of the AK is a function of how stable the decisions are, and we discuss a set of sustainability criteria and metrics useful to estimate the sustainability of this AK.

## What is Architectural Sustainability?

One of the success criteria for the architecture of a long-lived system is how well it survives the test of time in terms of supporting changes during the life cycle of the system while remaining intact. Many application domains, from complex engineering disciplines like automotive or avionics to information systems among others, demand stable architectures that are based on good, well-understood design decisions that extend the longevity of those architectures. However, in many situations software architects and developers struggle to cope with the impact of unpredictable changes (e.g., changes in technology platforms or surprising changes in the organization's business) that need extensive refactoring to implement them. We believe that this is often because architectural sustainability is not considered during the system design.

The longevity of a system often has a positive effect on the sustainability of the system, as these two factors can be seen as two sides of the software quality problem [1]. Architectural sustainability can be achieved through good design decisions that retain their validity over the long term, because such decisions have lasting value and influence [2]. From our perspective, we define architecture sustainability as "*the set of factors that influence the stability and longevity of an architecture during system evolution*". The notion of capturing architecturally "good" design decisions [3] can provide a basis for assessing the stability of such decisions over time if changes in the decisions do not affect the core of the resulting architecture. Consequently, there are three key questions for architects to ask concerning architecture sustainability:

(i) when is an architecture considered to be sustainable?
(ii) when can a decision be considered stable and what is the ideal lifetime of a design decision?
(iii) how can we measure the sustainability of architectural knowledge?

Given the number of long-lived systems in all domains today, these questions have become very relevant to all practicing software architects that want to raise or measure the quality, longevity, and stability of their architectures.

## Sustainable architectures

Estimating the sustainability of an architecture might not be easy, and we need to detect and identify "architecture smells" [4], showing that something is wrong or no longer adequate in the architecture [5]. These smells often arise as consequence of architecture-related technical debt[1] (TD), architectural mismatch or problems with the design decisions that have been made. Such loss of quality in the architecture usually affects the overall quality of the system, too; hence, we need metrics to estimate the ongoing quality of an architecture so that we can see when it starts to decay.

Another factor that clearly affects the architecture sustainability is how the evolution of the system is managed. In order to keep technical debt at acceptable levels, compensating technical changes must constantly be performed to repay the debt; otherwise, it will get out of control. We have found that there are a number of different categories of metrics that can be used to estimate the impact of changes on an architecture, which act as indicators of its sustainability:

- ripple effect metrics used to understand to what extent a change in a design decision affects other decisions (the higher the ripple effect, the poorer the architecture sustainability);
- instability, as opposed to stability, computed based on theory, as the probability of an architecture to change, while change proneness is computed empirically, and measures the effect of changes in architecture elements. Predicting instability or when a software module could change in a future version can be estimated as a probability function based on past changes and the percentage of ripple effects propagated from other modules; and
- code metrics used to detect anti-patterns in the architecture that consequently might lead to architecture changes. For instance, code metrics can provide indicators about complexity like coupling and cohesion and enable detecting, for instance, god classes that are clear signs of the Blob anti-pattern[2].

## Long-lived decisions

The sustainability of architectural knowledge can be achieved more easily if the knowledge is explicitly documented. Ideally this would be through the use of formally documented decision models, where the key design decisions could be captured and stored to be shared or even reused in new software projects. While formal models are rarely seen in industry, more lightweight, informal documentation can be used instead (e.g., capturing architectural knowledge in textual form perhaps using a template for guidance). Our experience leads us to strongly believe that the number of design decisions that have to be maintained (and the effort to maintain them) is a key indicator of the likely level of architecture sustainability of a system. This is primarily because the process of modifying a decision is not an isolated action and often influences other related decisions.

---

[1] http://martinfowler.com/bliki/TechnicalDebt.html
[2] https://sourcemaking.com/antipatterns/the-blob

As a result on our experience in this area, we base our assessment of the sustainability of an architecture on the following factors:

- the number of refactorings and frequency of changes performed over a period of time;
- the amount of significant design decisions changed; and
- the adequacy of the trace links between design decisions and other software artifacts that eases tracking when decisions are changed.

Good design decisions usually endure over a long period of time and enable the architecture to remain stable. Regarding the lifetime of good design decisions, these must be revisited in case of large architectural changes, but should not be constantly changed, as the system evolves. If we use appropriate metrics to monitor the size of the decision network, the number of changes to the decisions during evolution cycles, and the impact of refactorings, we can better estimate the extent to which an architecture can be considered to be sustainable. Therefore, the longevity of decisions is an indicator as to whether the architecture of a system is stable. Many contemporary decision model implementations contain a timestamp and a decision history to record when decisions are modified, which allows users to know the last time a decision changed. This is also useful when calculating metrics like the ones listed above and ones relating to the frequency of change.

We show in Figure 1 the factors that affect the sustainability and longevity of software architectures. Those factors derived from the complexity of a decision network or from the stability of decisions indicate some kind of technical debt, and we use these to identify a loss of quality in the architecture and hence, design erosion. The lower box in the figure represents different kinds of metrics and items that can be combined to estimate the sustainability of the architectural knowledge.

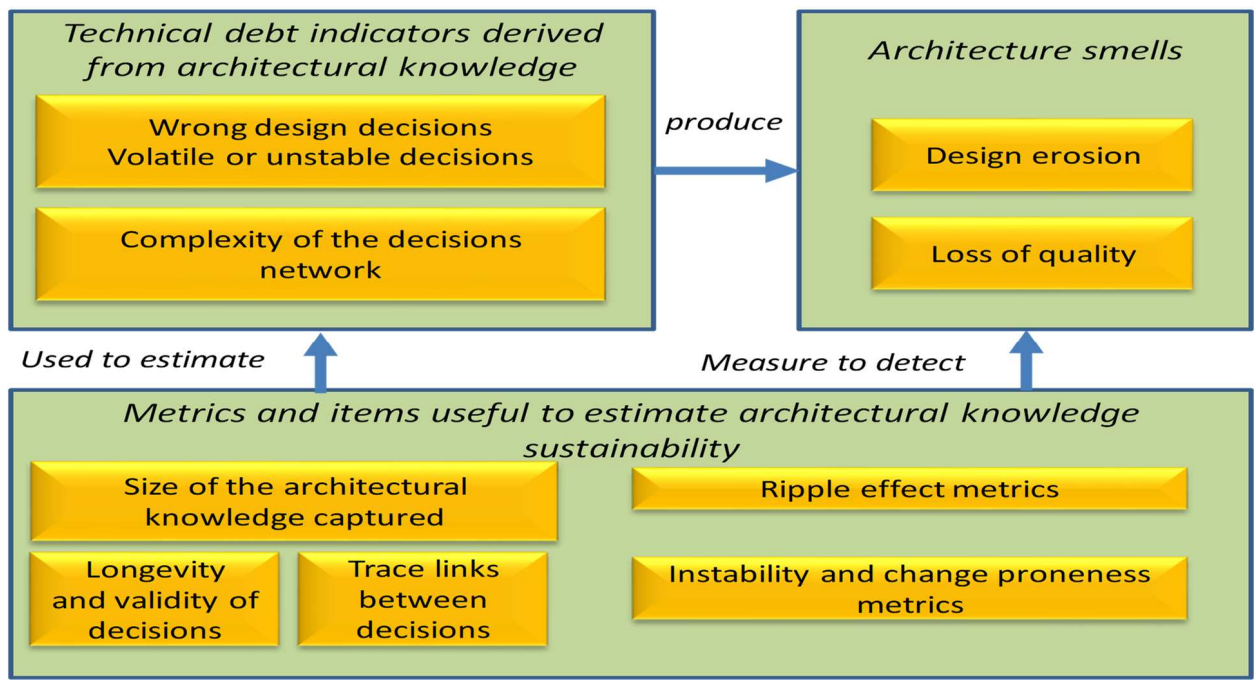

**Figure 1**: Architecture smells derived from metrics to estimate the sustainability of architectural knowledge.

## Measuring Sustainability of Architectural Knowledge

Because the majority of software metrics are code-oriented or design-oriented [6] [7], in a previous work [8] we suggested several criteria to estimate the sustainability of architectural knowledge, which we refine now in Table 1. The sustainability of architectural knowledge can not only be estimated in terms of how much effort we need to maintain the knowledge, but also how stable the decisions are and their longevity as the system evolves. Software maintainers can use this table as a guide to evaluate the sustainability of their architecture. It is important to both assess the values of the metrics at a specific point in and also to track trends in their values over time. We use the metrics in this table to measure the sustainability of the architectural knowledge and provide sustainability indicators for an architecture.

**Table 1**: Assessment criteria to measure architectural knowledge sustainability (based on [8]).

| Architectural Knowledge Practice Areas | Sustainability criteria for Architectural Knowledge | Quality attributes | Metrics |
|---|---|---|---|
| **Architectural Knowledge maintenance** Having a reduced set of design decisions and controlling the number of trace links between decisions and other software artifacts we ease the AK maintenance tasks when decisions change. | **Granularity of the design decisions and trace links.** This criterion limits the granularity of the decisions to be captured at the level of packages and classes, avoiding finer-grained decisions as a way to reduce the size of the decision model and make it more manageable. For example, decisions involving the creation of UML classes will be captured but not those concerning the creation of UML attributes or methods. Avoiding fine-grained decisions reduces the number of trace links to other software artifacts. | Complexity. The complexity of the decisions network can be reduced if we limit the granularity of the decisions captured and the number of trace links between them. Stability. Changes in the AK items do not affect to the decisions captured as no new nodes are needed in the decisions network when new AK items are added. | NodeCount EdgeCount Instability |
| | **Size of the decision model.** With this criterion we limit the number of design choices. Our experience in different projects suggests ranges of alternative decisions between [1:7/10]. | Cost of the effort capturing less number of decisions. | Number of Children |
| | **Number of AK attributes captured.** Capturing less amount of AK items using configurable AK templates makes the AK more manageable. We observed that capturing a number between 3 and 6 AK items seems reasonable. | Cost of the effort in capturing a number of variable AK items. | Number of Fields |
| **Architectural Knowledge evolution** Estimating better the number of decisions that will be impacted by a change and limiting this impact to a certain level helps to reduce the number of decisions to be analyzed during evolution cycles. Also, if we know in advance which decisions must be revisited we can predict better the stability and longevity of the AK model. | **Number of design decisions impacted.** Decisions that change have an impact on other related decisions. Limiting this impact using a ripple effect algorithm, we can reduce the number of decisions that must be analyzed or revised. However, the designer must establish this limit based on past experience. We can also limit the exploration to prune those decisions in the network that are "end-nodes". | Changeability: We can reduce the amount of effort to change the decisions impacted by a change limiting the number of decisions analyzed. Stability: If a decision changes less number of times, it affects to the stability of the architecture, as good decisions endure over time. | Change Impact Analysis (ripple effect) Change proneness Instability |
| | **Number of times a decision changes.** We can store information about how many times a decision changes in specific periods of times and use this information to control the stability of the decisions and how often they change. Therefore, we can provide indicators about the longevity of more stable decisions. | Stability: Decisions that change less often play a key role in favor of the stability of the architecture and longevity of the decisions as well. | Decision Volatility |
| | **Validity of decisions.** We can set specific dates for decision review and remove obsolete decisions that seem no longer valid. | Stability, Timeliness | Not yet defined but the timestamp of the decisions can be used |

## Using Metrics to Estimate the Sustainability of Architectural Knowledge

Table 1 suggests a set of metrics that software engineers can use to estimate the sustainability of architectural knowledge. We can combine several metrics to estimate a particular quality attribute. For instance, estimating the *complexity* of a decision network involves combining the "NodeCount", "EdgeCount", and "NumberOfChildren" metrics to estimate how complex, and hence how sustainable, the decision network is as it evolves. In those cases where we need to decrease the cost of the decisions captured and make the architectural knowledge capture more sustainable, the "NumberOfFields" indicator, combined with the number of decisions captured and the time spent in capturing them, serve as indicators to measure the ideal size of a decision model in different development contexts (e.g., agile versus RUP). For practitioners, we suggest the following steps to measure architectural knowledge sustainability using the criteria of Table 1:

- Select one quality attribute that impacts on the sustainability of the architectural knowledge;
- Review if the selected quality attribute matches one or several criteria for which they intend to measure the sustainability of the architectural knowledge;
- Define the input values that can be used to measure each criterion;
- Use algorithms from graph theory or code metrics that can be useful to measure the input values;
- Produce a normalized formula that can combine all the input values logically and use its result as an indicator of sustainability for the selected quality attribute; and
- Repeat the previous steps for all quality attributes relevant to estimate the sustainability of the architectural knowledge.

## Conclusion

Estimating the different aspects of architectural knowledge sustainability requires a number of different metrics, which when combined, can provide useful indicators about the health of an architecture. To achieve this, project teams need to have a good knowledge of the major architectural decisions in their projects and to capture this information. This can be done by simply documenting such decisions or by maintaining specific architectural knowledge documentation such as decision networks. When informal documentation is used, the metrics can be calculated manually and when formal documentation is used, automatic measurement is possible with the right tool support. To monitor the sustainability of the architectural knowledge, we can measure a range of the system's quality attributes. Meanwhile, emerging research is investigating the definition of new metrics to allow the sustainability of the design decisions to be computed automatically and reduce the effort required for architectural oversight. Finally, the criteria defined in Table 1 provide some practical guidance to help architects understand which items and metrics are useful to analyze the sustainability of their architectural knowledge.

**Acknowledgement:** We would like to gratefully thank Olaf Zimmermann and Eoin Woods for their thorough reviews and suggestions for improvement.

**RAFAEL CAPILLA** is Associate Professor at the Rey Juan Carlos University of Madrid. Contact him at rafael.capilla@urjc.es.

**ELISA YUMI NAKAGAWA** is Associate Professor at the University of São Paulo. Contact her at elisa@icmc.usp.br.

**UWE ZDUN** is Professor at the University of Vienna. Contact him at uwe.zdun@univie.ac.at.

**CARLOS CARRILLO** is Associate Professor at the Polytechnic University of Madrid. Contact him at carlos.carrillo@upm.es.